\documentclass[twocolumn,floatfix,prl,showpacs,superscriptaddress]{revtex4}
\usepackage{graphicx}
\usepackage{amsmath}
\usepackage{natbib}

\begin{document}
	\title{Pressure-tuning  the quantum spin Hamiltonian of the triangular lattice antiferromagnet Cs$_2$CuCl$_4$}

\author{S.~A.~Zvyagin}
\thanks{E-mail address: s.zvyagin@hzdr.de}
\affiliation{Dresden High Magnetic Field Laboratory (HLD-EMFL), Helmholtz-Zentrum Dresden-Rossendorf, 01328 Dresden, Germany}

\author{D.~Graf}
\affiliation{National High Magnetic Field Laboratory, Florida State University, Tallahassee, Florida 32310, USA}

\author{T.~Sakurai}
\affiliation{Research Facility Center for Science and Technology, Kobe University, Kobe 657-8501,
Japan}

\author{S.~Kimura}
\affiliation{Institute for Materials Research, Tohoku University, Sendai 980-8578, Japan}

\author{H.~Nojiri}
\affiliation{Institute for Materials Research, Tohoku University, Sendai 980-8578, Japan}

\author{J.~Wosnitza}
\affiliation{Dresden High Magnetic Field Laboratory (HLD-EMFL), Helmholtz-Zentrum Dresden-Rossendorf, 01328 Dresden, Germany}
\affiliation{Institut f\"{u}r Festk\"{o}rper- und Materialphysik, TU Dresden, 01062 Dresden, Germany}

\author{H.~Ohta}
\affiliation{Molecular Photoscience Research Center, Kobe University,  Kobe 657-8501,
Japan}

\author{T.~Ono}
\affiliation{Department of Physical Science, Osaka Prefecture University, Osaka 599-8531, Japan}

\author{H.~Tanaka}
\affiliation{Department of Physics, Tokyo Institute of Technology,  Tokyo 152-8551, Japan}

\date{\today}
	
\begin{abstract}

Quantum triangular-lattice antiferromagnets  are important prototype systems  to investigate phenomena of the geometrical frustration in condensed matter.  Apart from highly unusual magnetic properties,  they possess a rich  phase diagram (ranging from  an unfrustrated square lattice to a quantum spin liquid), yet to be confirmed experimentally. One major obstacle in this area of research is  the   lack of materials  with appropriate  (ideally tuned) magnetic parameters. Using Cs$_2$CuCl$_4$ as a model system, we demonstrate an alternative approach, where, instead of the chemical composition,  the spin Hamiltonian  is  altered by hydrostatic pressure. The approach combines high-pressure electron spin resonance and magnetization measurements, allowing us not only to quasi-continuously tune the exchange parameters,  but also  to accurately monitor them. Our experiments indicate a substantial  increase of the exchange  coupling ratio from 0.3 to 0.42 at a  pressure of 1.8 GPa, revealing a number of  emergent  field-induced phases.

\end{abstract}
	
\maketitle
	

\textbf{Introduction}

 The interplay between geometrical frustration, quantum fluctuations,  and magnetic order is one of the central issues in condensed matter  physics. In 1973, developing the resonating valence bond (RVB) theory, Anderson proposed that quantum fluctuations in magnetic structures on an isotropic  triangular  lattice can be sufficiently  strong to destroy the  magnetic order, resulting in a  two-dimensional (2D) fluid of mobile  spin pairs correlated together into singlets \cite{Anderson}. This state was introduced as a  RVB quantum spin liquid, contrary to the valence-bond solid (VBS), with the ground state condensed into a spin lattice. The  Anderson's hypothesis  has triggered  a cascade of extensive  theoretical  and experimental studies, resulting in the discovery of exotic  quantum states and  highly unusual field-induced phenomena \cite{Balents}.

\begin{figure} [!h]

\begin{center}
\vspace{0mm}
\includegraphics[width=0.5\textwidth]{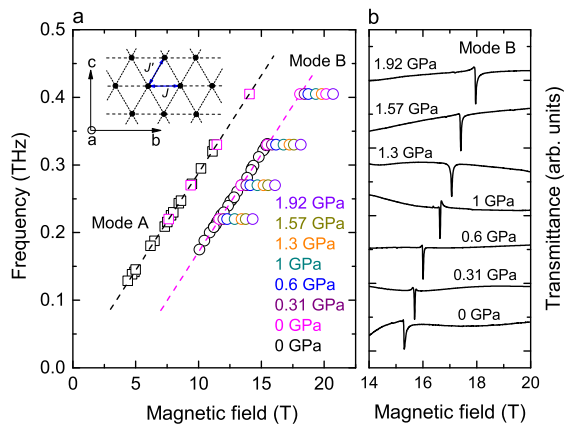}
\vspace{2mm}
\caption{\label{fig:ESR} Pressure dependence of ESR excitations in  Cs$_2$CuCl$_4$ ($T=1.9$ K, $H\parallel b$). 
(\textit{a}) Frequency-field diagrams of ESR  excitations  at different pressures. The data denoted in black  are taken from Ref. \cite{Zvyagin_SH} ($T=1.5$ K; 0 GPa). 
Dashed lines correspond to the fit results (see text for .details). The inset shows a schematic picture of magnetic sites and exchange couplings in a  triangular layer of  Cs$_2$CuCl$_4$. 
(\textit{b}) ESR spectra (mode B) taken at 330 GHz at different pressures (the spectra are offset for clarity).}
\end{center}
\end{figure}

The spin-1/2 triangular-lattice Heisenberg antiferromagnet (AF) represents one of the most important  classes  of the  family of low-D quantum frustrated magnets. For the general case of spatially anisotropic triangular AF the spin Hamiltonian is given as 
\begin{equation}
\mathcal{H} = J\sum_{\langle i,j \rangle} {\bf S}_i\cdot{\bf S}_j +
J' \sum_{\langle i,j' \rangle}  {\bf S}_i\cdot{\bf S}_{j'}, \
\label{Ham}
\end{equation}
where  ${\bf S}_i$, ${\bf S}_j$,  and ${\bf S}_{j'}$ are spin-1/2  operators at sites $i$, $j$, and $j'$,  and  $J$ and $J'$ are the exchange interactions on the horizontal and diagonal bonds,  respectively  (Fig.~\ref{fig:ESR}a, inset). In spite of  this  simple model,  such  systems  are shown to possess  a very rich and not fully understood  phase diagram, which can be interpolated between  decoupled spin-chain ($J'=0$),  isotropic triangular  ($J'/J=1$),  and  unfrustrated square ($J=0$) lattices. It is expected that transitions from one state to another occur in between these well-defined cases, but  many details of this evolution (e.g.,  critical coupling ratios) still remain a matter of debate \cite{Starykh_Rev,Chen}. The magnetic phase diagram   predicts a variety of exotic phases, with the 1/3 saturation-magnetization plateau as  the most exciting magnetic property \cite{Ono}.

The largest hindrances to experimentally check  theoretical predictions on the unusual magnetic properties of  spin-1/2 triangular lattice Heisenberg AFs  is the very limited number of  materials with  appropriate  (ideally  tuned) sets  of  parameters, currently available for measurements. In spite of the recent progress in synthesizing numerous  spin-1/2 triangular-lattice materials (see, e.g.,  \cite{Balents} and references therein),  the two compounds, Cs$_2$CuCl$_4$ and Cs$_2$CuBr$_4$ (with $J'/J \simeq 0.30$ and  0.41, respectively  \cite{Zvyagin_SH}),   remain among the most prominent   representatives of this family of frustrated materials.  One obvious approach to tune the  spin Hamiltonian  of these systems is to vary   their chemical composition \cite{Ono_DOP,Well}. However,  experiments on the solid solution  Cs$_2$CuCl$_{4-x}$Br$_x$  (with Br content ranging from 0 to 4)  revealed a pronounced difference  in  the  Cu  coordination when increasing  $x$, resulting in  a discontinuous  evolution  of  its  crystal structure \cite{Cong}.

 The high-pressure  technique is known as a powerful means  to  modify magnetic properties and  parameters of  exchange coupled spin systems  (see e.g. \cite{Zal,Goto,Ruegg,Hong,Graf,Graf_2,Komu,Zayed,Skou,Weh}.    On the other hand, another important task  is to precisely measure these parameters. This becomes particular challenging  for low-D spin systems, whose spin Hamiltonian is strongly affected by quantum fluctuations.  One solution to solve this problem   is to suppress quantum fluctuations by strong-enough magnetic fields, and then to use the harmonic spin-wave theory for description of the excitation spectrum \cite{Zvyagin_SH,Coldea}. 

 Our approach combines high-pressure high-field    electron spin resonance (ESR) and magnetization measurements, allowing us not only to quasi-continuously change the exchange parameters $J$ and $J'$, but also to accurately monitor them.  We use  Cs$_2$CuCl$_4$ as a model system.   We showed  that the application of  pressure increases significantly the exchange coupling parameters in this compound, triggering, at the same time, the emergence of  field-induced low-temperature magnetic phases, absent at zero pressure.

\textbf{Results} 

\textbf{High-pressure ESR measurements.} To  determine  the dependence of the  coupling parameters of Cs$_2$CuCl$_4$ on the applied pressure, we used the procedure  employed  in Ref. \cite{Zvyagin_SH}, when  the  excitation  spectrum is measured above the saturation field  $H_{sat}$. In the case of the staggered  Dzyaloshinskii-Moriya (DM) interaction the ESR  spectrum should consist of two modes,  which correspond to magnetic excitations at the center and at the boundary of the unfolded Brillouin zone (a.k.a. the relativistic and exchange modes, respectively).  Such modes were previously  observed  in Cs$_2$CuCl$_4$ \cite{Zvyagin_SH} (black symbols  in Fig.~\ref{fig:ESR}a). 
The field dependence  of the relativistic mode A  for $H \gtrsim J/g\mu_B$ can be  described using the equation $\hbar\omega_A = g\mu_B H$, where $\hbar$ is the reduced Planck constant, $\omega$ is the excitation frequency, $\mu_B$ is the Bohr magneton, and  $g=2.06$ is the $g$ factor (the fit results are shown in Fig.~\ref{fig:ESR}a by the black dashed line). On the other hand, the frequency-field diagram of mode B  can be described using the equation $\hbar \omega_B =  g\mu_B H - \Delta_B$ (magenta dashed line in Fig.~\ref{fig:ESR}a) with the same $g$ factor  as for the mode A.   Most importantly,  the difference between the excitation energies  for modes A and B ($\Delta \omega_{AB} \equiv \Delta_B$) is determined by $J'$: $J'=\hbar\Delta \omega_{AB}/4$, allowing us to measure $J'$ directly.

\begin{figure} [!h]

\begin{center}
\vspace{0mm}
\includegraphics[width=0.5\textwidth]{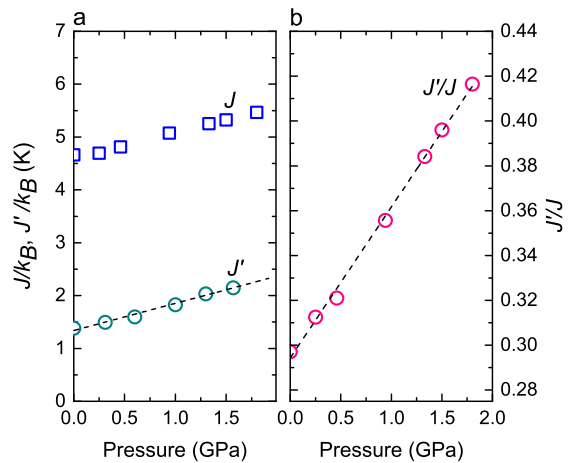}
\vspace{-3mm}
\caption{\label{fig:JJ}  Pressure-driven tuning of  the spin-Hamiltonian parameters in Cs$_2$CuCl$_4$.  (\textit{a}) Pressure dependence of the exchange coupling parameters $J'$ and $J$  (circles and boxes, respectively). The dashed line corresponds to  a  linear fit  to the $J'$ data (see text for details).  (\textit{b}) Pressure dependence of the exchange coupling ratio $J'/J$. The dashed line corresponds to  a  linear fit  to the $J'/J$ data (see text for  details).}
\end{center}
\end{figure}

The experiment revealed a shift of the mode B  towards higher field when the pressure is applied. The pressure dependence of  $J'$ is  shown in Fig.~\ref{fig:JJ}a,  evident  in a significant, almost 70$\%$,   increase of $J'$ at  1.92 GPa.

\begin{figure} [!h]

\begin{center}
\vspace{-0mm}
\includegraphics[width=0.5\textwidth]{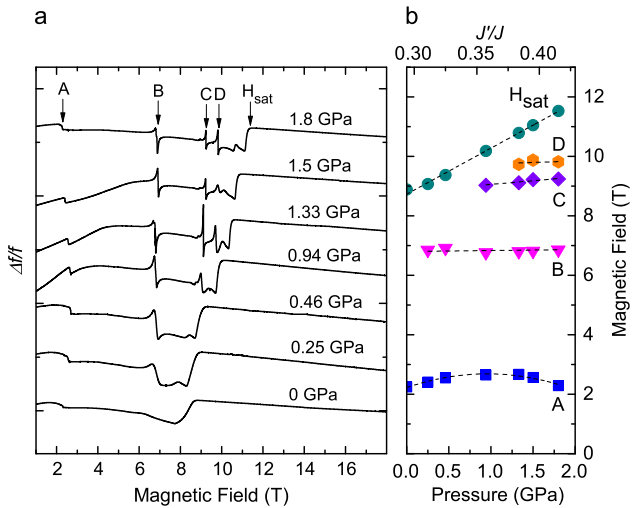}
\vspace{-3mm}
\caption{\label{fig:TDO}  Pressure evolution of  magnetic properties of Cs$_2$CuCl$_4$ obtained by means of TDO technique  ($T=350$ mK, $H\parallel b$).    (\textit{a}) Pressure dependence of  the TDO  frequency changes $\Delta f/f$ in response to magnetic field  (the data are offset for clarity). (\textit{b}) Dependencies of TDO frequency anomalies on the applied pressure. The  calculated exchange  coupling ratio $J'/J$ is shown on the top scale (see text for the details). Lines are guides for the eye. Source data are provided as a Source Data file.}
\end{center}
\end{figure}

\textbf{High-pressure TDO measurements.}  Knowing $J'$ and the  saturation field $H_{sat}$, we can determine $J$,  using the expression $g\mu_B H_{sat} = 2J(1+J'/2J)^2$.
 To measure the saturation field of Cs$_2$CuCl$_4$  we employ a  radio frequency tunnel-diode-oscillator (TDO) technique  (see Methods).    The variations of the TDO circuit resonant frequency $\Delta f/f$ as a function of magnetic field  applied along the $b$ axis at different pressures  are shown in Fig.~\ref{fig:TDO}a. In strong magnetic fields, the  TDO frequency is almost constant,  indicating  the transition of Cs$_2$CuCl$_4$ into  the fully spin-polarized  phase with saturated magnetization  \cite{Tokiwa}.  The experiment revealed that  with  increasing pressure the saturation field moves toward higher magnetic fields. The dependence of $H_{sat}$ on the applied pressure is shown in  Fig.~\ref{fig:TDO}b.

Based on the combined ESR and TDO magnetization data, for zero pressure we obtained $J'/k_B= 1.38$ K and $J/k_B= 4.66$~K ($J'/J \simeq 0.3$), which perfectly agrees with previous estimates \cite{Zvyagin_SH}.   Results of a linear  fit  to the  $J'$ dependence (dashed line in Fig.~\ref{fig:JJ}a) were used to calculate $J$ at different pressures.  $J'$, $J$ and $J'/J$ as functions of the applied pressure are shown in Fig.~\ref{fig:JJ}.  The $J'/J$ dependence can be described  using the empirical  equation $J'/J=0.294(2) + 0.067(2)\cdot P$ (dashed line in Fig.~\ref{fig:JJ}b), where $P$ is the applied pressure (GPa).  For  1.8 GPa,  we obtained  $J'/k_B = 2.28$ K, $J/k_B = 5.47$ K, and   $J'/J \simeq 0.42$,   indicating a remarkable, by 40 $\%$, increase of the $J'/J$ ratio. Based on this fit,   the application  of a pressure of 3.6 GPa (where Cs$_2$CuCl$_4$  undergoes a structural phase transition \cite{Xu}) would allow one to  reach   $J'/J \simeq 0.53$ (which corresponds to  approximately  180\%  of the zero-pressure value).

\textbf{Discussion}

Apart from the  shift of the saturation field,  our experiment revealed a number of magnetic anomalies, which are absent in Cs$_2$CuCl$_4$ at zero pressure (Fig.~\ref{fig:TDO}a). 
The  observed magnetic anomalies can be caused by  changes in the  dynamics of critical fluctuations  in the vicinity of  field-induced phase transitions \cite{Kawasaki}, resulting in changes of  real and imaginary components of the  magnetic susceptibility. Although no signature of the  1/3 magnetization plateau was revealed, our observation  (Fig.~\ref{fig:TDO}b) resembles   the  cascade of field-induced phase  transitions  in quasi-2D  Cs$_2$CuBr$_4$ \cite{Fortune},  evident of a complex picture of magnetic interactions in both materials (a remarkable  sensitivity of the  magnetic phase diagrams of Cs$_2$CuCl$_4$ to the direction of the applied  magnetic field \cite{Tokiwa} strongly suggests  an important role not only spatial ($J' \neq J$), but also spin-space (asymmetric DM interaction) components of the magnetic anisotropy; the latter  appear to be of the same order of magnitude as the interplane exchange interaction $J''$ \cite{Coldea},  inducing  strongly relevant perturbations  \cite{Starykh_ES}). 

 For the magnetic field applied along the $b$ axis the zero-pressure magnetic phase diagram contains  four low-temperature phases \cite{Tokiwa}.  At small field below $T_N=0.62$ K,  the system is in the incommensurate phase with a spiral ground state \cite{Coldea_NS}  dominantly determined by the  DM anisotropy (``DM spiral'') \cite{Starykh_ES}. In this phase the spins are located almost in the $b$-$c$ plane with  the spiral propagating   along the $b$ axis \cite{Coldea_NS}. Remarkably, at about 2.3 T the effect of the DM interaction becomes  irrelevant and the system  undergoes a transition into the commensurate coplanar AF phase with  spins  more correlated in $a$-$b$ planes   (the corresponding correlations are determined by  $J''$ and $J$)  \cite{Starykh_ES}. These two magnetic phases are  stabilized by quantum fluctuations.  The commensurate coplanar AF state is realized in a relatively wide field range, followed by two  successive high-field transitions:    into the  noncoplanar cone phase and then, with further increase of the applied magnetic field,    into the fully spin polarized magnetically saturated phase (both phases are favored classically).

\begin{figure} [!h]

\begin{center}
\vspace{-10mm}
\includegraphics[width=0.55\textwidth]{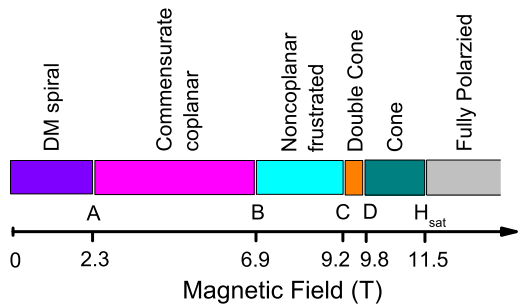}
\vspace{-25mm}
\caption{\label{fig:DIAGR}  The proposed phase diagram  of  Cs$_2$CuCl$_4$ under pressure 1.8 GPa (which corresponds to $J'/J=0.42$).  }
\end{center}
\end{figure}

What happens when pressure is applied? Apart from the  shift of the saturation field,  our experiment revealed a number of magnetic transitions, absent at zero pressure  (Fig.~\ref{fig:TDO}a). The  proposed magnetic phase diagram for 1.8 GPa is shown in Fig.~\ref{fig:DIAGR}. Similar to that at zero pressure, at low field the system is in the DM spiral phase.  The DM spiral phase  is suppressed by magnetic field at  about 2.2 - 2.6  T (the anomaly A in  Fig.~\ref{fig:TDO}a corresponds to this transition),  resulting in the  commensurate coplanar AF phase with spins predominantly correlated in the $a$-$b$ plane.  Applied  pressure makes  the  $J'$ term more and more relevant,  tending  to suppress the coplanar nature of magnetic correlations.  As a combined  effect  of the applied magnetic field (partially suppressing quantum order) and pressure (enhancing the interplane correlations),  at about 6.9 T the system undergoes  a transition into a noncoplanar  frustrated phase.  The observed anomaly B corresponds to this  transition.

For a spatially anisotropic triangular lattice AF in  magnetic fields near the saturation,   theory  \cite{Starykh_Jin} predicts  a particular rich phase diagram,  with ground states ranging from an 
incommensurate noncoplanar chiral cone    to a commensurate coplanar $V$ state.  The transformation between these two states  involves two intermediate phases. One of them is a coplanar incommensurate  order, while another one is  a noncoplanar double-\textbf{Q} spiral order (double-cone state). The latter is characterized by the broken  $Z_2$ symmetry between  two magnon condensates at $\pm Q$ (where $Q$ is the ordering wave vector) and can coexist with the single-cone phase in a relatively narrow range of  $J'/J$, but at smaller  fields.  In Cs$_2$CuCl$_4$ at zero pressure, the transition into the single-cone  phase  was revealed  between 8 and 9 T  below  300 mK \cite{Tokiwa}. Due to the  increase of  exchange coupling parameters, the applied pressure shifts the  upper boundary of the temperature-field phase diagram to higher temperatures. Because of that the transition into the single-cone phase  can be observed  at higher  temperatures.  Based on this assumption, the anomalies C and D can be interpreted as  transitions into the double- and single-cone phases, respectively  (Fig.~\ref{fig:TDO}).  A tiny feature  immediately before  saturation  might indicate  the involvement of other higher-order perturbation factors (e.g., next-nearest-neighbor interactions \cite{Ye}  or  the interplane frustration mentioned above \cite{Batista}).

Our observations call for systematic high-pressure magneto-structural (nuclear magnetic resonance and neutron diffraction)  studies of Cs$_2$CuCl$_4$, which would allow one to verify  the proposed phase diagram.  Apart from exact identification of the nature of the observed  high-pressure phases, another important task  would be the search for the  field-induced  1/3 magnetization plateau,   which can be  expected with  further increase of $J'/J$ moving the system towards the  isotropic ($J'/J=1$) limit \cite{Starykh_Jin}. It would be also very interesting to measure the pressure-driven evolution of the spin Hamiltonian in  the isostructural compound Cs$_2$CuBr$_4$ and to compare the results with that in  Cs$_2$CuCl$_4$.

To conclude, we demonstrated  an effective strategy to control the spin Hamiltonian of a spin-1/2 antiferromagnet  on a triangular lattice with  hydrostatic pressure. With increasing pressure,  for  Cs$_2$CuCl$_4$ our experiments  revealed  a substantial  increase of the exchange  coupling parameters, accompanied by the emergence of (at least)  two field-induced phases. These phases can be tentatively  interpreted as noncoplanar frustrated and double-cone states, merging the low-field commensurate coplanar and high-field single-cone phases revealed previously. Our  approach provides robust means for investigating  the complex interplay between geometrical frustration, quantum fluctuations,  and magnetic order (especially, close to quantum phase transitions), paving  the  way  towards  controlled  manipulation of the spin Hamiltonian and magnetic properties  of frustrated spin systems.

\textbf{Methods} 

\textbf{Single-crystal growth.} Single-crystal samples of Cs$_2$CuCl$_4$ were  grown by the slow evaporation of aqueous solution of CsCl and CuCl$_2$ in the mole ratio $2:1$.  

\textbf{High-pressure TDO.} High-pressure TDO measurements   were conducted  at the  National High Magnetic Field Laboratory (Florida State University) in magnetic fields up to 18 T using  a TDO  magnetometer  \cite{Clover,Graf,Graf_2}   tuned to operate at a resonant frequency of ~ 51 MHz. Magnetic field was applied along the $b$ axis of the crystal.  A sample  with a length of $\sim 1.5$ mm  was placed in a  copper-wire coil  with diameter $\sim 0.8$ mm and height $\sim1$ mm. The coil and sample were surrounded with Daphne 7575 oil (Idemitsu Kosan Co., Ltd.) and encapsulated in a Teflon cup which was inserted into the bore of a piston-cylinder pressure cell  constructed from a chromium alloy (MP35N). The coil acts as an inductor in a diode-biased self-resonant LC tank circuit. 
During the field sweep,  changes in the sample  magnetic permeability (i.e. dynamic magnetic susceptibility)   lead to changes in the inductance of the oscillator tank coil, and, hence, to changes in the TDO circuit resonant frequency $\Delta f$.  The frequency changes were detected as a function of magnetic field  at different pressures. 
The pressure created in the cell was calibrated at room temperature and again at low temperature using the fluorescence of the R1 peak of a small ruby chip as a pressure marker \cite{Piermarini} with accuracy better than $\pm 0.015$ GPa.  The pressure cell was immersed directly into $^3$He, allowing TDO measurements down to 350 mK. Temperature was measured using a calibrated Cernox thermometer. Transition fields were measured with accuracy better than $\pm 0.5\%$.

\textbf{High-pressure ESR.} 
High-pressure ESR measurements of Cs$_2$CuCl$_4$  were performed at the High Field Laboratory for Superconducting Materials,   Institute for Material Research (IMR), Tohoku University using  a transmission-type ESR probe \cite{Sakurai,Sakurai_2} with oversized waveguides and a 25 T cryogen-free superconducting magnet  \cite{Awaji,Sakurai_3}.   Gunn-oscillators, operated at frequencies 220, 270, 330, and 405 GHz, were employed  as  radiation sources. A hot-electron InSb bolometer cooled down to 4.2 K  was used as  detector.    Magnetic field was applied along the $b$ axis of the crystal. Experiments were performed at a temperature of 1.9(1) K; the temperature was measured using a calibrated Cernox thermometer. A cylinder-shaped crystal  with approximate dimensions of 9 mm in length  by 5 mm in diameter was immersed in a Teflon cup filled with Daphne 7474 oil (Idemitsu Kosan Co., Ltd.) as  pressure medium. A two-section   piston cylinder pressure cell made from NiCrAl (inner cylinder)  and CuBe (outer sleeve) has been used. The key feature of the pressure cell is the inner pistons, made of ZrO$_2$  ceramics; this material has low loss for electromagnetic radiation  with frequency up to 800 GHz.  The change of the superconducting transition temperature of tin was used to calibrate the applied pressure \cite{Smith}; the transition temperature was detected by AC magnetic susceptibility measurements.  Applied pressure was calculated using  the relation between the load at room temperature and the pressure obtained at around 3 K \cite{Sakurai}; the pressure calibration accuracy is  better than $\pm 0.02$ GPa.  ESR line position was measured with accuracy better than  $\pm 0.2\%$. In our experiments we assume that the accuracies  estimating $J'$ and $J$, including all possible error sources,  are better than $\pm 1\%$ and $\pm 4\%$, respectively.

\textbf{Data availability}

The data that support the findings of this study are available from the corresponding 
author upon reasonable request. The source data underlying Figs 1a, 2a,b, and 3b are provided as  Source Data files.

\textbf{Author contributions}

S.Z. conceived, designed and led the project.  T.O. and H.T. grew Cs$_2$CuCl$_4$ single crystals. D.G. and S.Z. performed high-field magnetization experiments. T.S., S.K., H.N., and S.Z.  performed high-field ESR experiments.  J.W. and H. O. administered the HLD and KU parts of the project, respectively. All authors discussed the results and commented on the manuscript.

\textbf{Acknowledgment}

This work was partially supported by the Deutsche Forschungsgemeinschaft (projects ZV6/2-2, SFB 1143, and through the W\"{u}rzburg-Dresden Cluster of Excellence on Complexity and Topology in Quantum Matter - $ct.qmat$ (EXC 2147, project-id 39085490)). We acknowledge support by  the HLD at HZDR, member of the European Magnetic Field Laboratory (EMFL). ESR experiments were performed at the High Field Laboratory for Superconducting Materials,   Institute for Material Research, Tohoku University (proposals 18H0059  and 17H0071). H.N. acknowledges the support of the KAKENHI 16H04005 Program. S.Z. acknowledges the support of the ICC-IMR Visitor Program at the Tohoku University. A portion of this work was performed at the National High Magnetic Field Laboratory, which is supported by the National Science Foundation Cooperative Agreement No. DMR-1644779 and the State of Florida.  Authors and would like to thank  O. A. ~Starykh, R.~Valent\'{\i}, C.D. ~Batista,  L.~Balents and R. Coldea for  fruitful discussions, and A. ~Ponomaryov for the help orienting the samples.

\textbf{Additional information}

The authors declare no competing financial or non-financial interests.

\end{document}